\documentclass{article}

\usepackage[margin=1in]{geometry}
\usepackage{svg}
\usepackage{lipsum}
\usepackage{amsmath}
\usepackage{amsfonts}
\begin{document}

\title{\textbf{Terahertz-induced cascaded interactions between spectra offset by large frequencies}}
\author{Koustuban Ravi $^{1,2,*}$, and Franz X. K\"artner $^{1,2,3}$}
\maketitle
\date{}
\noindent

\begin{center}
\textit{1. Center for Free-Electron Laser Science, DESY,Notkestra$\beta$e 85, Hamburg 22607, Germany\\
2. Research Laboratory of Electronics, Massachusetts Institute of Technology\\
3. Department of Physics, University of Hamburg, Hamburg 22761, Germany\\
e-mail : koustuban.ravi@cfel.de}
\end{center}

\textbf {Abstract}: We explore the dynamics of a system where input spectra in the optical domain with very disparate center frequencies are strongly coupled via highly phase-matched, cascaded second-order nonlinear processes driven by terahertz radiation. The only requirement is that one of the input spectra contain sufficient bandwidth to generate the phase-matched terahertz-frequency driver. The frequency separation between the input spectra can be more than ten times larger than the phase-matched terahertz frequency. A practical application of such a system where the cascading of a narrowband pump line centered at 1064\,nm induced by a group of weaker seed lines centered about 1030\,nm and separated by the phase-matched terahertz frequency is introduced. This approach is predicted to generate terahertz radiation with percent-level conversion efficiencies and millijoule-level pulse energies in cryogenically-cooled periodically poled lithium niobate. A model that solves for the nonlinear coupled interaction of terahertz and optical waves is employed. The calculations account for second and third-order nonlinearities, dispersion in the optical and terahertz domains as well as terahertz absorption. Ramifications of pulse formats on laser-induced damage are estimated by tracking the generated free-electron density. Strategies to mitigate laser-induced damage are also outlined.

\section{Introduction}
Nonlinear optical interactions which give rise to further coupled nonlinear interactions in a \textit{cascade} of processes maybe broadly referred to as cascaded optical nonlinearities. A few examples are found in \cite{schiek1996one, varan1997, Golomb04, moses06, bache2007scaling, bache2017cascaded}.

In many cases, cascaded nonlinearities are not all, well phase matched \cite{varan1997, bache2017cascaded}. However, consider the case of difference-frequency generation between near-degenerate waves in the optical or near-infrared (NIR) spectral range yielding terahertz radiation. Here, a large number of highly phase-matched three-wave mixing processes occur in concert. This is because terahertz frequencies are much smaller than optical frequencies, which enables many phase-matched processes to be simultaneously supported. Therefore, very interesting spectral dynamics are observed in such systems.

Cascading effects in terahertz generation systems have predominantly received attention in the context of optical-to-terahertz conversion efficiencies beyond the single-photon conversion or Manley-Rowe limit \cite{Golomb04,vodo2006,vodo2008,jewariya09}. In such cases, cascaded processes result in the repeated energy down-conversion of a single optical (or NIR) photon to produce multiple terahertz photons \cite{Golomb04}. 

However, cascading effects in terahertz generation systems can be important even when absolute optical-to-terahertz conversion efficiencies are low. The simplest example of such a system is one with optical parametric amplifier (OPA) like initial conditions. This involves a pump in the optical or NIR region and a signal/idler in the terahertz region. Several experiments in various configurations yielding relatively low-to-modest conversion efficiencies along with modulated optical spectra have been reported in this regard \cite{walsh2010intracavity,Molter09,Tripathi:14,murate2018perspective, cirmi2017,Hemmer18}. 

It was theoretically shown that even for $\approx 0 \%$ optical-to-terahertz energy conversion efficiency, such configurations can only be reasonably described by considering cascading effects, i.e. the simultaneous evolution of a large number of coupled three-wave mixing processes \cite{ravi16}. 

\begin{figure}
\centering
\includegraphics[scale=0.5]{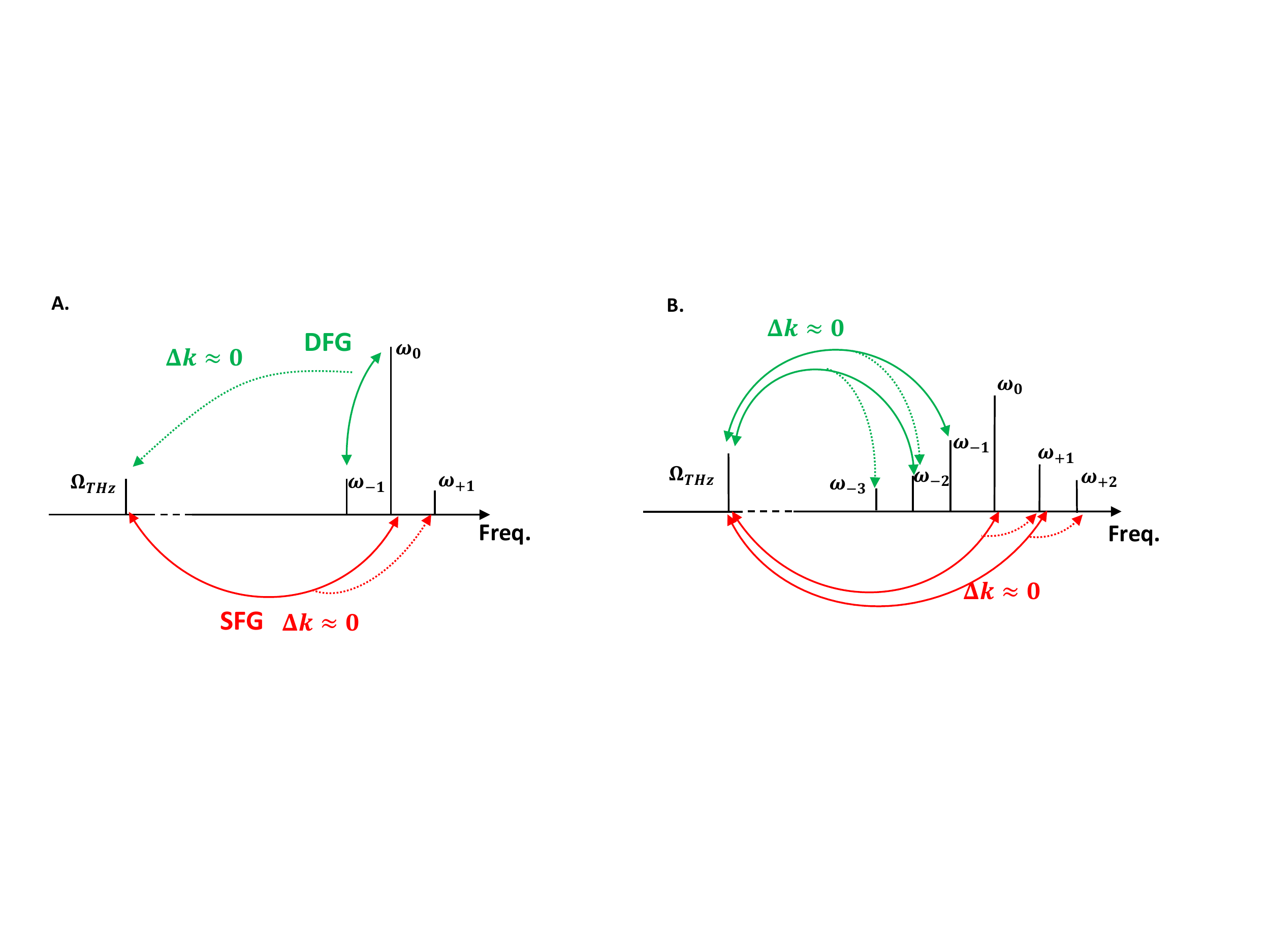}
\caption{\label{fig_1} Example of a previously studied terahertz generation system where it is necessary to consider the simultaneous evolution of many phase-matched second-order processes even if optical-to-terahertz energy conversion efficiencies are low. (a) Initial interactions in a cascaded terahertz parametric amplifier. An OPA-like input of a strong optical pump at $\omega_0$ and weaker optical seed at $\omega_{-1}$ is used. The double arrows delineate the pair of waves involved in the interaction and the dotted arrows indicate the product of those interactions. For an input comprised of a strong pump at $\omega_0$ and weak seed at $\omega_{-1}$ in the optical domain, a terahertz wave at $\Omega_{THz}$ corresponding to the beat frequency is first generated which then drives via sum frequency generation, the component at $\omega_{+1}$. (b) In subsequent steps, the generated terahertz radiation drives generation of red-shifted components $\omega_{-2,-3}$ and further blue-shifted components, e.g. $\omega_{+2}$. SFG consumes a terahertz photon while DFG generates one. Therefore, absolute conversion efficiencies are low in the absence of a mechanism which prefers DFG.}
\end{figure}

The reasons for this are apparent by inspecting the schematic in Fig. \ref{fig_1}, which describes the spectral dynamics of a terahertz generation system with OPA-like initial conditions. Consider the initial condition when a strong optical pump with angular frequency $\omega_0$ and optical seed of weaker intensity with angular frequency $\omega_{-1}=\omega_0-\Omega_{THz}$ are phase matched (i.e. $\Delta k=0$) for the difference-frequency generation (DFG) of terahertz radiation with angular frequency $\Omega_{THz}$. The initial interactions are depicted in Fig. \ref{fig_1}(a). Here, the line at $\omega_0$ beats with $\omega_{-1}$ to generate $\Omega_{THz}$, which in turn beats with $\omega_0$ to generate $\omega_{-1}$. Mathematically, the phase-mismatch $\Delta k=k(\omega_0)-k(\omega_0-\Omega_{THz})-k(\Omega_{THz})=0$ (where $k$ is the wave number). However, since $\Omega_{THz}\ll\omega_0$, $k(\omega_0+\Omega_{THz})-k(\omega_0)-k(\Omega_{THz})$ is also approximately zero. Therefore, the sum-frequency generation (SFG) of $\omega_{+1}=\omega_0+\Omega_{THz}$ is also phase matched. Each side-band $\omega_{\pm 1}$ can undergo further  phase-matched SFG and DFG processes to produce a series of lines separated by $\Omega_{THz}$ as shown in Fig. \ref{fig_1}(b). Each SFG process consumes a terahertz photon while each DFG process creates one. Therefore, in the absence of a mechanism to preferentially phase-match DFG processes, only modulation of terahertz radiation will occur. At the same time, the optical spectrum would exhibit many side-bands \cite{Molter09,cirmi2017,Hemmer18}. This illustrates the importance of considering the simultaneous evolution of many strongly coupled nonlinear interactions or cascading effects in a system with OPA-like initial conditions.

The consideration of cascading effects for OPA-like inputs predict many other anomalies \cite{ravi16}. For instance, the wave of highest intensity could be present at a lower frequency ($\omega_{-1}$) to produce terahertz amplification, which is disallowed in conventional parametric amplifiers. This is because the input optical spectrum is rapidly washed out due to the simultaneous evolution of many strongly coupled phase-matched processes and hence becomes inconsequential.

Building on this understanding of the nature of strongly coupled phase-matched processes in terahertz generation systems, in the present work we explore the spectral dynamics of a much more sophisticated system. It is comprised of input optical/NIR spectra separated by a very large frequency offset. However, one of the spectra contains sufficient bandwidth to generate phase-matched terahertz radiation. The generated terahertz radiation, then drives cascaded interactions with the other spectrum (See Section 2, Fig. \ref{fig_2} for an illustration). 

The above format lends itself to useful practical applications. The first involves coupling a single narrowband line of high energy at 1064\,nm to a set of much weaker seed lines centered at 1030\,nm. The seed lines are separated by the phase-matched sub-terahertz frequency. It is worthwhile to note that the separation between pump and seed is about 10\,THz, which is more than ten times larger than the generated terahertz frequency. Furthermore, note that the higher energy pump line is situated at a lower frequency in relation to the lower energy seed lines. 

The terahertz radiation generated by beating of lines centered about 1030\,nm, drives the cascading of the line at 1064\,nm. Eventually, the two spectra merge and subsequently an effective red-shift of the total optical spectrum results to yield very efficient terahertz generation at the percent-level in cryogenically-cooled periodically poled lithium niobate (PPLN). The availability of high energy Nd:YAG lasers at 1064\,nm makes this approach particularly relevant to high energy terahertz generation. For pump fluences on the order of $0.5$\,Jcm$^{-2}$ in conjunction with cm$^2$-aperture crystals \cite{ishizuki2014}, these conversion efficiencies translate to millijoule-level terahertz pulse energies with hundreds of cycles. Such pulses are of interest for compact particle acceleration \cite{nanni2015,zhang2018,lemery2018} and X-ray free-electron lasers \cite{kartner2016}. 

In section 2, we outline the general physics of the approach. In section 3, we present detailed numerical studies describing the physics of the system. The effect of the input spectrum on laser-induced damage is studied and methods to mitigate it are suggested. In section 4, we present numerical results of terahertz generation using a 1064\,nm pump and a set of seed lines centered at 1030\,nm. We conclude in section 5.

\section{Physical mechanism}
\begin{figure}
\centering
\includegraphics[scale=0.5]{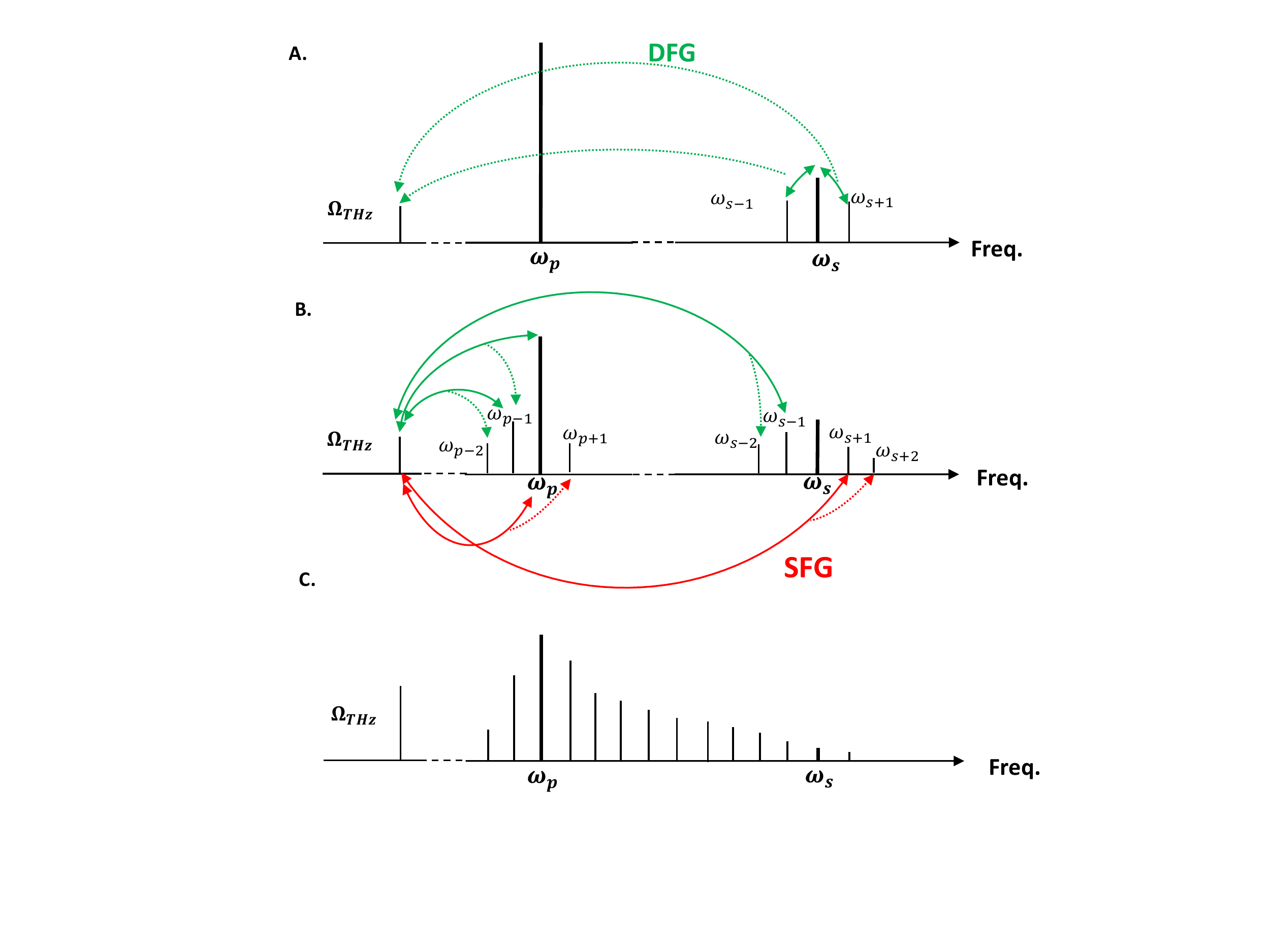}
\caption{\label{fig_2} Schematic describing the spectral dynamics of the present system. (a) A pump (higher intensity) and seed (lower intensity) with very disparate center frequencies. The angular frequency of the pump $\omega_p$ may be larger or smaller than $\omega_s$. However, a few lines separated by the terahertz angular frequency $\Omega_{THz}$ are centered about the seed at $\omega_s$, which beat to generate terahertz radiation. This subsequently mediates the cascaded processes. (b) The generated terahertz radiation first modulates both the pump and seed lines, producing a series of side-bands $\omega_{s\pm m}$ and $\omega_{p\pm m}$ ($m=1,2,\dots$), about the seed and pump respectively by DFG and SFG processes. (c) Eventually, the spectral gap between the pump and seed reduces and the process produces a net red-shift if phase matching conditions are favorable. This is accompanied with an increase in terahertz generation.}
\end{figure}

In Fig. \ref{fig_2}, a schematic illustrating the cascading dynamics governing the interaction between two input optical spectra separated by a frequency much larger than the phase-matched terahertz frequency is depicted.

\begin{figure}
    \centering
    \includegraphics[scale=0.55]{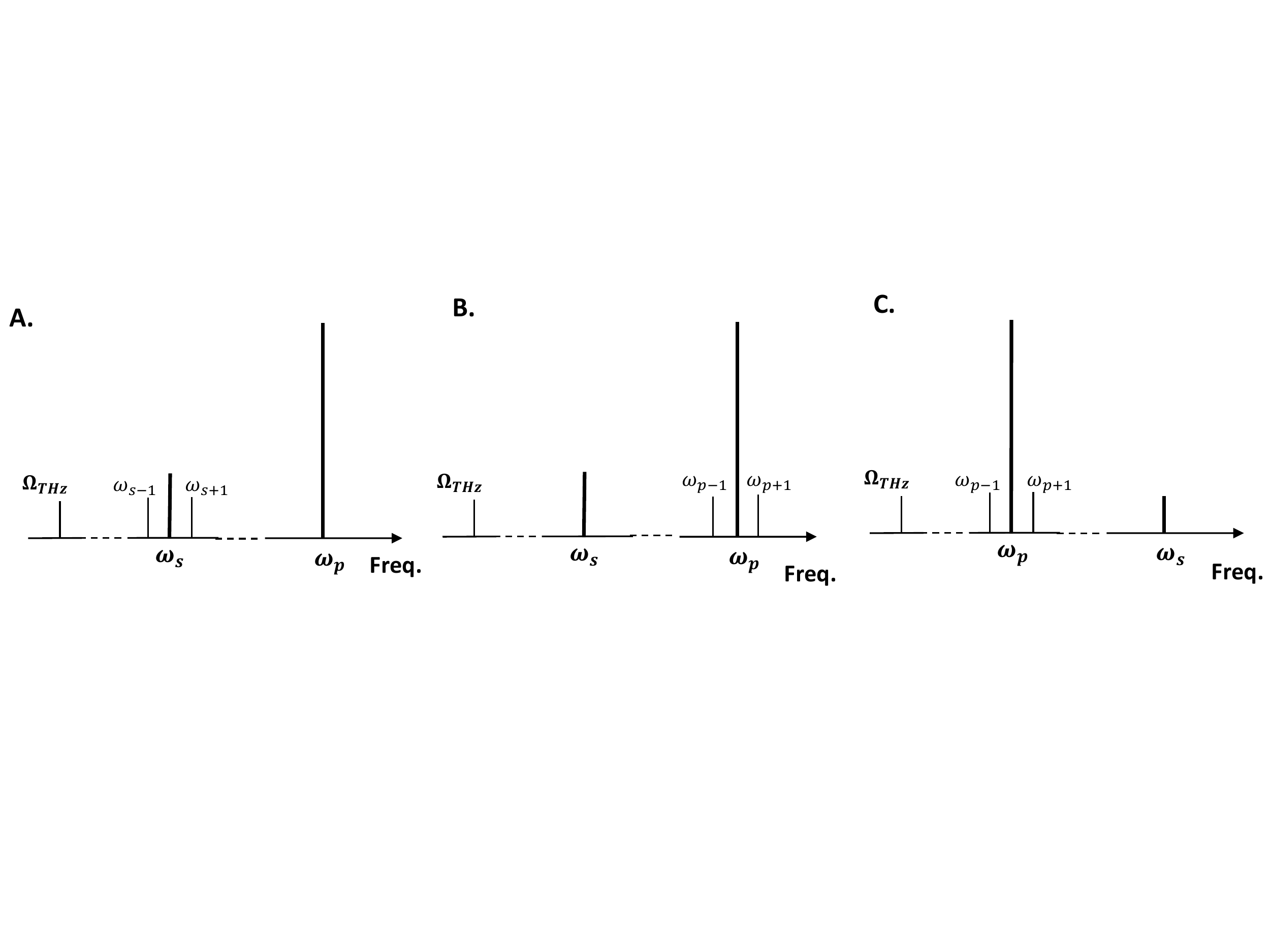}
    \caption{Variations in input spectral formats producing similar dynamics. (a) Seed is located at a smaller angular frequency, i.e. $\omega_s<\omega_p$. (b) Additional lines to generate terahertz radiation are distributed about the pump rather than the seed, i.e. $\omega_{p\pm1}=\omega_p\pm\Omega_{THz}$ and $\omega_p<\omega_s$.(c) Additional lines are provided adjacent to the pump and $\omega_p>\omega_s$. In all cases, the cascading dynamics will be similar if phase-matching towards red-shifted frequencies is superior.}
    \label{pf_var}
\end{figure}

The pump spectrum is narrowband, centered at angular frequency $\omega_p$ and is represented by a single line in Fig. \ref{fig_2}(a). Experimentally, this spectral distribution may correspond to a long pulse with a transform-limited duration of $>100\,$ps. The seed is centered at angular frequency $\omega_s$ and contains a set of lines separated by the phase-matched terahertz angular frequency $\Omega_{THz}$. A minimum of two lines shall be required for the seed. Each seed line may correspond to a narrowband pulse, also of transform-limited duration $> 100\,$ps. Variants of this input pulse format are described at the end of this section and depicted in Fig. \ref{pf_var}. The physical processes at work for those formats can be readily deduced based on the discussion in this paper.

Here and throughout this paper, the pump is defined to be the pulse (or line) with larger intensity. The angular frequency of the pump $\omega_p$ may be at a smaller or larger frequency compared to the seed $\omega_s$. In terahertz generation systems, one of the peculiar consequences of cascading effects is that the initial conditions tend to rapidly wash out due to the simultaneous evolution of many strongly coupled three-wave mixing processes. This renders the relative location of pump and seed unimportant.

In the initial step as shown in Fig. \ref{fig_2}(b), DFG between spectral components centered around $\omega_s$ produces some terahertz radiation at $\Omega_{THz}$. Subsequently, the generated terahertz radiation interacts with the pump at $\omega_p$ and $\omega_s$ to produce side-bands $\omega_{p\pm1},\omega_{p\pm2}\dots$ and $\omega_{s\pm 1}, \omega_{s\pm 2}\dots$ about both frequencies respectively. As previously mentioned, this is because both SFG and DFG processes are well phase matched due to the small value of $\Omega_{THz}$ compared to $\omega_p$ or $\omega_s$.

If phase matching conditions for frequencies red-shifted with respect to the pump, i.e. for $\omega <\omega_p$ are more favorable,  a preferential red-shift of the \textit{total} (or combined) optical spectrum eventually occurs as shown in the final schematic in Fig. \ref{fig_2}(c). This red-shift is accompanied with significant terahertz generation. 

For the envisaged dynamics in Figs. \ref{fig_2}(a)-(c), terahertz absorption would have to be low enough to enable sufficient interaction. In practice, this can be realized in cryogenically-cooled lithium niobate.

In subsequent sections, we will numerically prove that the spectral evolution follows the above description and furnish further details on the process.

In addition to the input spectral format described in Fig. \ref{fig_2}(a), other variations depicted in Fig. \ref{pf_var} can also be utilized. In Fig. \ref{pf_var}(a), the seed is situated at an angular frequency smaller than the pump, i.e. $\omega_s<\omega_p$. In Fig. \ref{pf_var}(b), the additional lines to generate terahertz radiation are distributed about $\omega_p$ rather than $\omega_s$. Finally, in Fig. \ref{pf_var}(c), $\omega_p<\omega_s$ and contains additional lines distributed about $\omega_p$. The cascading dynamics in all these cases are similar to Fig. \ref{fig_2} and require phase-matching to be superior in the direction of  red-shift for efficient terahertz generation.

\section{Results}

\subsection{Numerical model}
\begin{table}[]
    \centering
    \begin{tabular}{c c c}
    \hline\\
    Parameter & Symbol &Value\\
    \hline\\
    Second-order nonlinear susceptibility&$\chi^{(2)}_0$&336\,pm/V\\
    Refractive index (Optical/NIR)& $n(\omega_m)$&\cite{jundt1997}\\
    Terahertz-frequency dependent refractive index & $n_{THz}(\Omega_{THz})$&\cite{fulop2011}\\
    Terahertz-frequency dependent absorption coefficient & $\alpha (\Omega_{THz})$ &\cite{palfalvi2005}\\
    Energy bandgap & $E_g$ & 4\,eV\\
    Avalanche coefficient & $g_{av}$ & $1.2\times10^{-4}$\,m$^{2}$J$^{-1}$ \cite{meng2015} \\
    Four-photon absorption coefficient & $\beta^{(4)}$& $1.7\times10^{-44}\,$m$^{5}$W$^{-3}$ \cite{Hoffmann07}\\
    Laser-induced damage threshold fluence & $F_d$& $1(\tau/100\text{ps})^{1/2}\,$Jcm$^{-2}$ \cite{ravi2016_ps}\\
    Nonlinear refractive index&$n_2$&1.25$\times 10^{-19}$m$^2$/W\cite{ravi2016_ps}\\
    \end{tabular}
    \caption{Parameters used in simulations}
    \label{tab1}
\end{table}

To study the general physics governing the approach, a model which approximates each spectral line as being monochromatic is used. In reality, each line has a narrow bandwidth corresponding to a transform-limited duration of $>100\,$ps. In the final section, we model the complete continuous spectral distribution without any approximations as in \cite{ravi2016_ps} for reliable quantitative predictions. The discrete and continuous approaches are in qualitative as well as reasonable quantitative agreement, which justifies the discrete approximation. The terahertz spectral envelope as a function of longitudinal distance $z$ is represented by $A_{THz}(z)$, while the optical/NIR ones are represented by $A_m(z)$. Their evolution is presented in Eqs. (\ref{thz}) and (\ref{op}) respectively.

\begin{subequations}
\begin{gather}
\frac{ d A_{THz}(z)}{dz} = -\frac{\alpha}{2}A_{THz}(z)
-\frac{j\Omega^2\chi^{(2)}(z)}{2k(\Omega)c^2}\sum_m A_{m+1}(z)A_{m}(z)^{*}e^{-j[k_{m+1}-k_m-k(\Omega)]z}\label{thz}\\
\frac{d A_{m}(z)}{dz} = \frac{-j\omega_m^2\chi^{(2)}(z)}{2k_m c^2}\bigg[A_{m+1}(z)A_{THz}^{*}(z)e^{-j[k_{m+1}-k_m-k(\Omega)]z}+A_{m-1}(z)A_{THz}(z)e^{-j[k_{m-1}+k(\Omega)-k_m]z}\bigg]\label{op}
\end{gather}
\end{subequations}

In Eq. (\ref{thz}), the first term represents terahertz absorption while the second represents the ensemble of DFG processes between all spectral components in the optical domain. The second-order nonlinear susceptibility is denoted by $\chi^{(2)}(z)$, with the $z$ dependence provided to account for poled electro-optic crystals. In Eq. (\ref{op}), the first term within square brackets represents beating of the $m+1^{th}$ spectral component in the optical domain and terahertz radiation to generate the spectral component $m$. The second term however represents SFG of the $m^{th}$ spectral component in the optical domain by addition of the frequencies of the $m-1^{th}$  spectral component and terahertz radiation. 

To provide a quantitative illustration of the picture outlined in the previous section, we consider seed lines with an intensity distribution $I_{s,m}$ given by Eq. \ref{seed_dist}.

\begin{gather}
I_{s,m}= I_s\textbf{exp}[-(\omega_m-2\pi f_c)^2/(N_w f_{THz})^2] \label{seed_dist}
\end{gather}

In Eq. (\ref{seed_dist}), $f_c$ is the spectral separation between the two input spectra (pump and seed) while $N_w$ represents the bandwidth of the seed distribution in integer units (or number of lines). $I_s$ denotes the peak intensity among all seed lines. In the current simulation, $f_c=10$\,THz and $N_w=2$. The terahertz frequency $f_{THz}=\Omega_{THz}/(2\pi)$ is set to be 0.5\,THz. A single pump line is considered with $I_s= 0.01I_p$. 

A PPLN crystal with 5$\%$ Magnesium Oxide (MgO) doping is assumed to be phase matched for the generation of $f_{THz}=0.5\,$THz under conditions of cryogenic cooling at $T=100\,$K. Full material dispersion in the optical domain is obtained from \cite{jundt1997}, while that for the terahertz region is obtained from \cite{palfalvi2005}. Terahertz absorption coefficients at cryogenic temperatures are obtained from \cite{fulop2011}. Using these material properties, the required poling period is obtained as $\Lambda = 240\,\mu$m. The second-order nonlinear susceptibility of bulk lithium niobate is assumed to be $\chi^{(2)}_0=336\,$pm/V. The various material parameters used in this paper are tabulated in Table \ref{tab1}.

Each line is assumed to correspond to a transform-limited $e^{-2}$ duration of $\tau=200\,$ps. The total fluence, i.e. for all pump and seed lines put together is set to $0.7\,$Jcm$^{-2}$. This corresponds to half the laser-induced damage threshold value, which is given by the relation $F_{d} = 1(\tau/100\text{ps})^{1/2}\,\text{J}/\text{cm}^{-2}$ \cite{ravi2016_ps}. 

\begin{figure}
\centering
\includegraphics[scale=0.35]{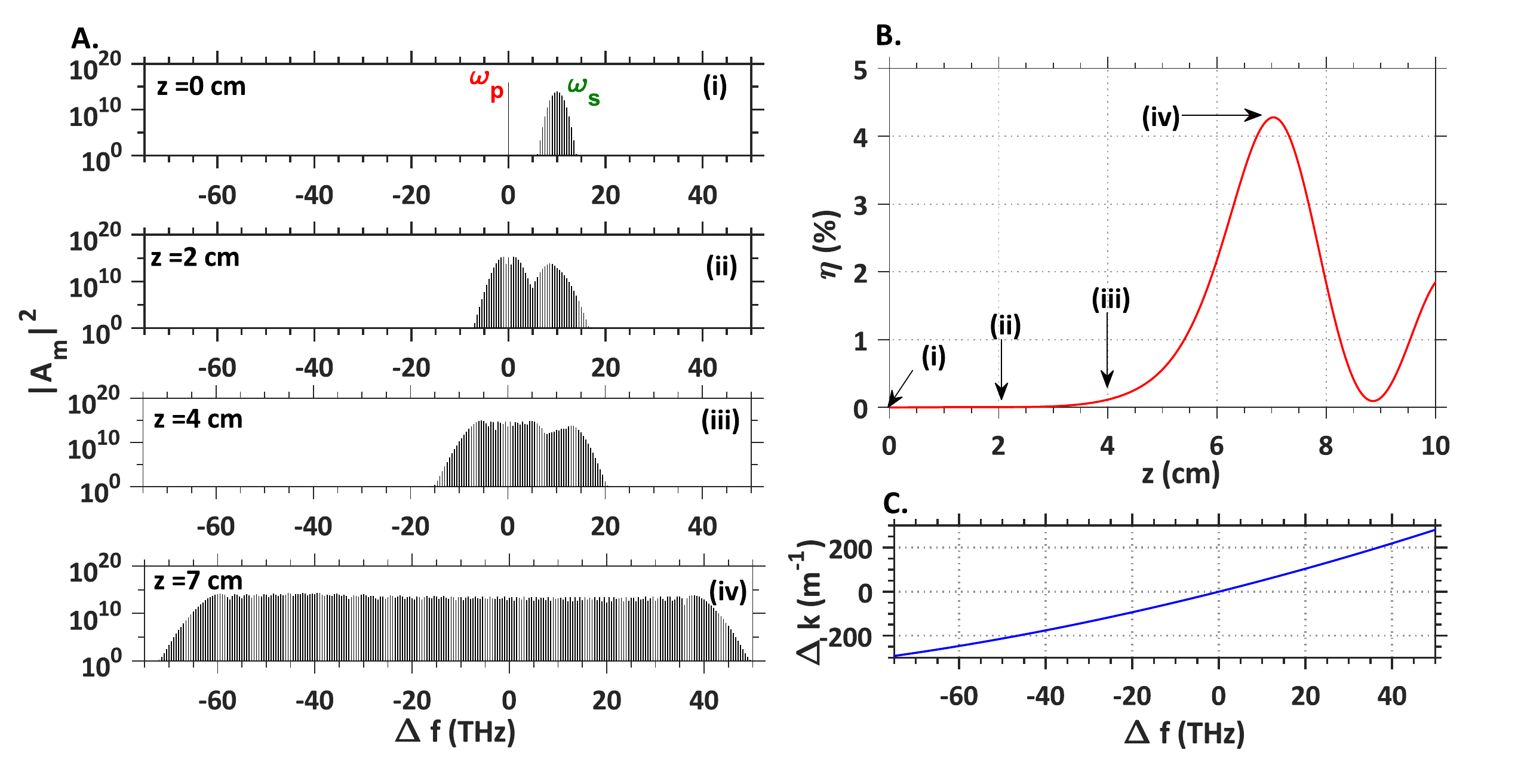}
\caption{\label{fig3} (a) The \textit{total} optical spectrum (i.e. pump and seed) is plotted for various locations along the crystal $z$ in panels (i)-(iv). Initially, the pump and seed spectra are very disparate in panel (i). After initial modulation by the generated terahertz radiation, they begin to merge as seen at location (ii). Subsequently, they completely merge in (iii), followed by significant preferential red-shift in panel (iv). The merging of the spectra, i.e. at z=4\,cm is seen to be the point at which conversion efficiency begins to grow rapidly. (b) Terahertz efficiency growth as a function of propagation length $z$ for $f_c=10$\,THz, $I_s=0.01I_p$ and $N_w=2$ in cryogenically-cooled PPLN crystals. An exponential growth of terahertz conversion efficiency, followed by a drop due to phase mismatch is observed. (c) The phase-mismatch as a function of detuning from the signal frequency is plotted. Phase-matching for red shifting with respect to $\omega_p$ is preferred.}
\end{figure}

In Fig. \ref{fig3}(a), the total optical spectrum, i.e. of both pump and seed are depicted. The panels labeled (i)-(iv) delineate the total spectrum at different locations along the propagation direction $z$ inside the crystal. 

As can be seen, the initial input format consists of two spectra centered at angular frequencies $\omega_s$ and $\omega_p$. The spectral distribution about $\omega_s$ is of significantly lower intensity and contains lines separated by the terahertz frequency $f_{THz}$. Initially, the generated terahertz radiation is weak and functions primarily as a means to modulate the two input spectra, resulting in a gradual merging of the pump and seed spectra as can be seen in panel (ii) of Fig. \ref{fig3}(a). Only after $z=4$\,cm, does the modulation give way to a preferential red-shift as can be seen in panels (iii)-(iv) of Fig. \ref{fig3}(a).

In Fig. \ref{fig3}(b), we plot the terahertz conversion efficiency as a function of length. The arrows labeled (i)-(iv) correspond to the locations represented by panels (i)-(iv) in Fig. \ref{fig3}(a).
No terahertz radiation is incident on the crystal and therefore the conversion efficiency $\eta$ at $z=0$ is zero. The conversion efficiency remains low at $z=2\,$cm, since the generated terahertz radiation has been primarily modulating the pump and seed spectra as evident in panel (ii) of Fig. \ref{fig3}(a). It is only at $z\approx4\,$cm that $\eta$ begins to experience significant growth. This coincides with a preferential red-shift in the total spectrum as is seen in panel (iii) of Fig. \ref{fig3}(a). The terahertz conversion efficiency then grows exponentially over the crystal length, followed by a drop beyond $z\approx7\,$cm. This drop is due to phase mismatch which occurs as the total optical bandwidth increases by spectral broadening. 

Figure \ref{fig3}(c) depicts phase matching as a function of detuning from the pump frequency. As can be seen, phase mismatch is larger for the region of blue shift ($\Delta f>0$) compared to the region of red shift ($\Delta f<0$). This shall be shown to be a key requirement for growth in terahertz efficiency.

\subsection{Effects of dispersion}

To further understand the cascading process in this system, we examine its behavior under various conditions of dispersion. Absorption is switched off to not obfuscate the overarching physics governing the process.  

\begin{figure}
\centering
\includegraphics[scale=0.35]{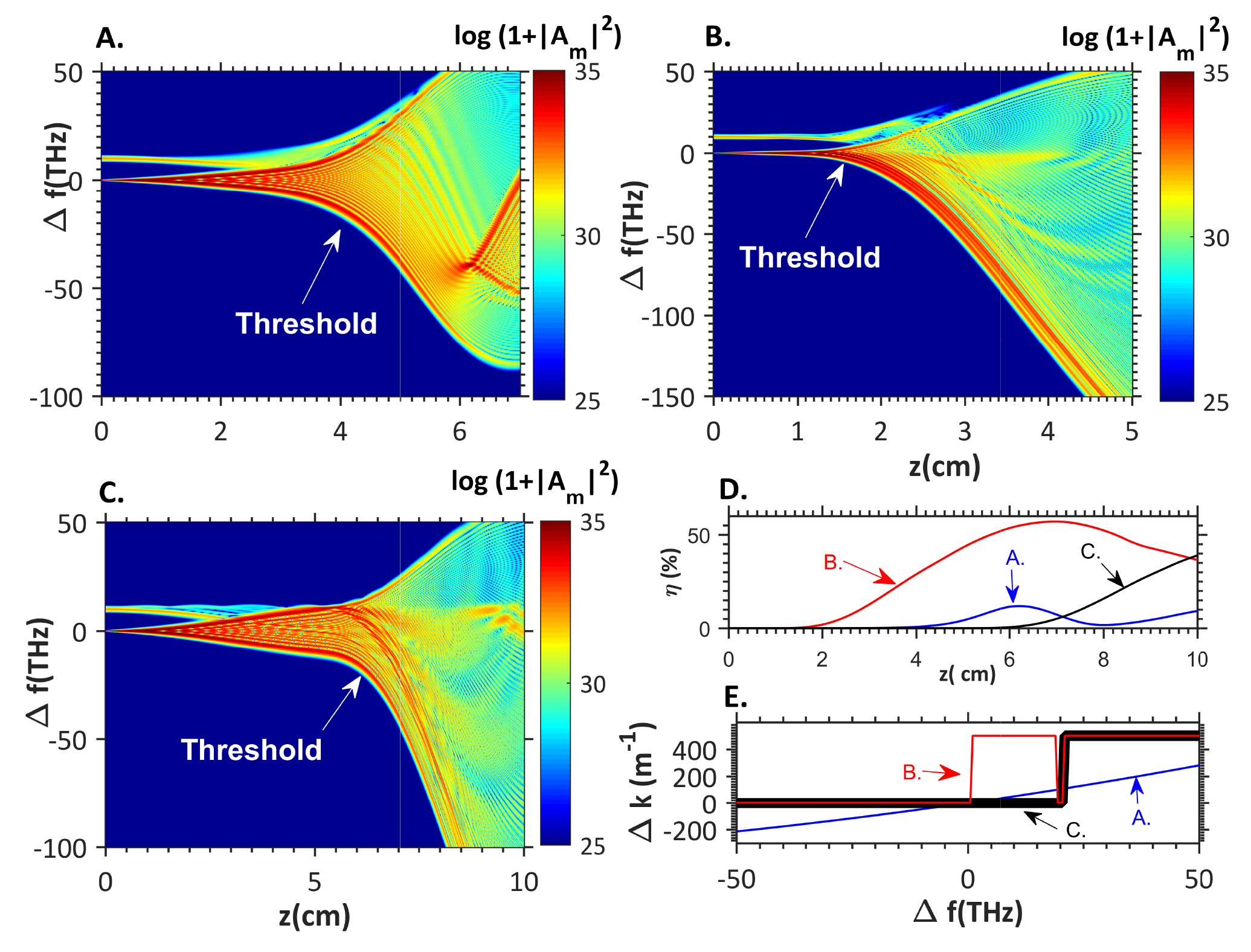}
\caption{\label{fig4}Dispersion curves and their influence on cascading dynamics. Absorption is switched off to not obfuscate the overarching physics without loss of generality. (a) Plot showing evolution of the total optical spectrum along crystal length for conventional phase-matching conditions in PPLN. This represents the experimentally relevant case. Initial modulation of pump and seed spectra around $\omega_p,\omega_s$ respectively is evident. Effective red-shift occurs only when the two spectra merge at $\approx 4\,$cm. This is followed by preferential red-shift and then subsequently back-conversion due to phase-mismatch. (b) Cascading dynamics for a fictitious dispersion curve where $\omega_p<\omega<\omega_s$ and $\omega>\omega_s$ are very highly phase mismatched. In this situation, the initially generated terahertz radiation by beating between lines distributed about $\omega_s$, drives the continuous red-shift of the spectrum about $\omega_p$. Since the need for modulation is greatly reduced, significant red-shift occurs at much shorter distances of $z=2\,$cm. (c) A dispersion curve, where only $\omega>\omega_s$ is phase-mismatched. In this case, the spectrum about $\omega_p$ will blue shift to a much larger extent till it reaches $\omega_s$ before commencing a red shift. Thus, the threshold point occurs at distances $z=6\,$cm, i.e. larger than that in (a). (d) Conversion efficiencies for various cases (a)-(c). (e) Dispersion in phase matching corresponding to the cases in (a)-(c).}
\end{figure}

In Fig. \ref{fig4}(a), the total optical spectrum as a function of propagation distance $z$ is plotted for the case when dispersion in phase matching is identical to that depicted in the previous section (blue curve in Fig. \ref{fig4}(e)). This represents the experimentally relevant case. In Fig. \ref{fig4}(b), the spectral dynamics for a situation where $\Delta k =0$ except for $\omega_p<\omega<\omega_s$ and for $\omega>\omega_s$ (red curve in Fig. \ref{fig4}(e)) is shown. Finally, in Fig. \ref{fig4}(c), $\Delta k \neq 0$ only for $\omega>\omega_s$ (black, Fig. \ref{fig4}(e)).

In Fig. \ref{fig4}(a), the terahertz radiation generated by DFG between lines distributed around the seed centered at $\omega_s$ initially modulates both pump and seed spectra until preferential phase matching in the region red-shifted with respect to $\omega_s$, i.e. $\Delta f<0$, begins to take precedence over blue-shifting. This occurs at $z\approx4\,$cm, which is when optical-to-terahertz conversion efficiency begins to grow as is evident in the blue curve in Fig. \ref{fig4}(d).

In Fig. \ref{fig4}(b), the region $\omega_p<\omega<\omega_s$ is so highly phase mismatched that effective red shifting occurs right from the outset, leading to much earlier growth in terahertz efficiency at $z\approx2\,$cm (red curve, Fig. \ref{fig4}(d)). This is because, blue-shifting with respect to the seed is prohibited by the phase mismatch in that region.

In Fig. \ref{fig4}(c), phase matching is violated only for $\omega>\omega_s$. Therefore, the pump continues to be modulated until it spreads from $\omega_p$ till it reaches the spectral distribution around $\omega_s$. At this point, when blue shifting can no longer occur, the red-shifted processes begin to dominate and terahertz efficiency growth occurs. Since the spectrum has to spread to a much greater extent compared to cases (a) and (b), the growth of terahertz conversion efficiency occurs much later at $z \approx 6\,$cm as can be seen in Fig. \ref{fig4}(d) (black curve).

Thus from previous numerical simulations, it can be seen that the main requirement for a net red-shift is that phase matching must improve for red-shifted optical/NIR frequencies. A caveat however is that the distance along the propagation direction $z$ at which preferential red shift occurs must be much larger than the terahertz absorption length. Otherwise, the terahertz field would be too strongly attenuated before the modulated spectrum reaches a point of large phase mismatch (in the blue-shifted region) to result in preferential red shift.

\subsection{Effect of $N_w$}

\begin{figure}
\centering
\includegraphics[scale=0.4]{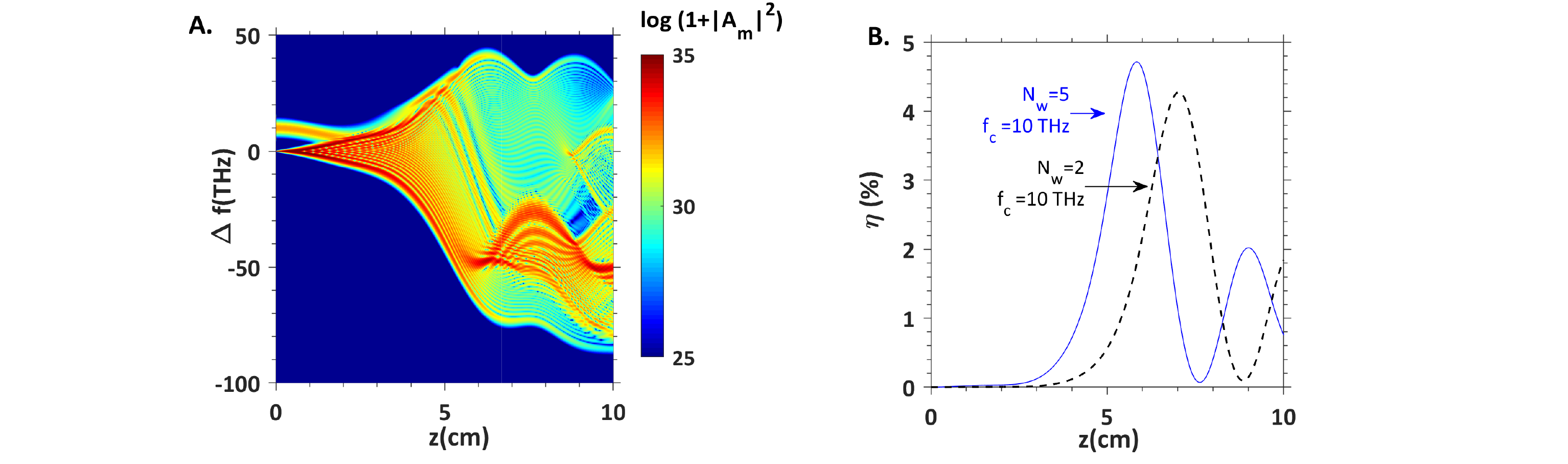}
\caption{\label{fig5}(a) Spectral dynamics for $N_w=5 $ lines about $\omega_s$ separated by $f_c= 10$\,THz from $\omega_p$. (b) The threshold point at which terahertz efficiency experiences growth is reduced compared to the case of $N_w=2,f_c=10\,$THz.}
\end{figure}

In this section, we investigate the effect of the bandwidth of the seed $N_w$ (or loosely the number of seed lines) on the cascading dynamics in Fig. \ref{fig5}. The role of the frequency separation $f_c$ between the pump and seed spectra can be similarly deduced. In the ensuing simulations, the exact material parameters for lithium niobate at $T=100\,$K from Table \ref{tab1} are assumed. A PPLN with $\Lambda=240\,\mu$m, corresponding to a phase-matched terahertz frequency of $0.5\,$THz is considered.

In general, increasing $N_w$ should result in the faster onset of preferential red shift. This is because the distance at which the two input spectra are modulated to an extent that they begin to overlap and force a net red shift decreases. This can be seen by comparing Fig. \ref{fig5}(a) to Fig. \ref{fig4}(a). In Fig. \ref{fig5}(a), $f_c=10\,$THz as in Fig. \ref{fig4}(a) but $N_w$ is increased to 5 as opposed to $N_w=2$ in Fig. \ref{fig4}(a). 

This is further evident upon examining the terahertz conversion efficiency as a function of propagation distance $z$ in Fig. \ref{fig5}(b). Here, the cases from  Fig. \ref{fig5}(a) ($N_w=5,\,f_c=10\,$THz) experiences exponential growth at a shorter distance of $z=3\,$cm compared to the case when $N_w=2,\,f_c=10\,$THz (black-dotted curve in Fig. \ref{fig5}(b)).

Decreasing $f_c$ would have a similar effect to increasing $N_w$ since it too would result in an effective reduction of the gap between the pump and seed spectra.

From the above set of numerical experiments in Figs. \ref{fig3}-\ref{fig5}, we have deduced that phase matching for frequencies lower than the pump must be better in the case when $\omega_p<\omega_s$. Furthermore, we find that the onset of efficiency growth occurs only when the pump and seed spectra begin to overlap. 

\subsection{Effect of terahertz frequency and seed intensity}

\begin{figure}
\centering
\includegraphics[scale=0.45]{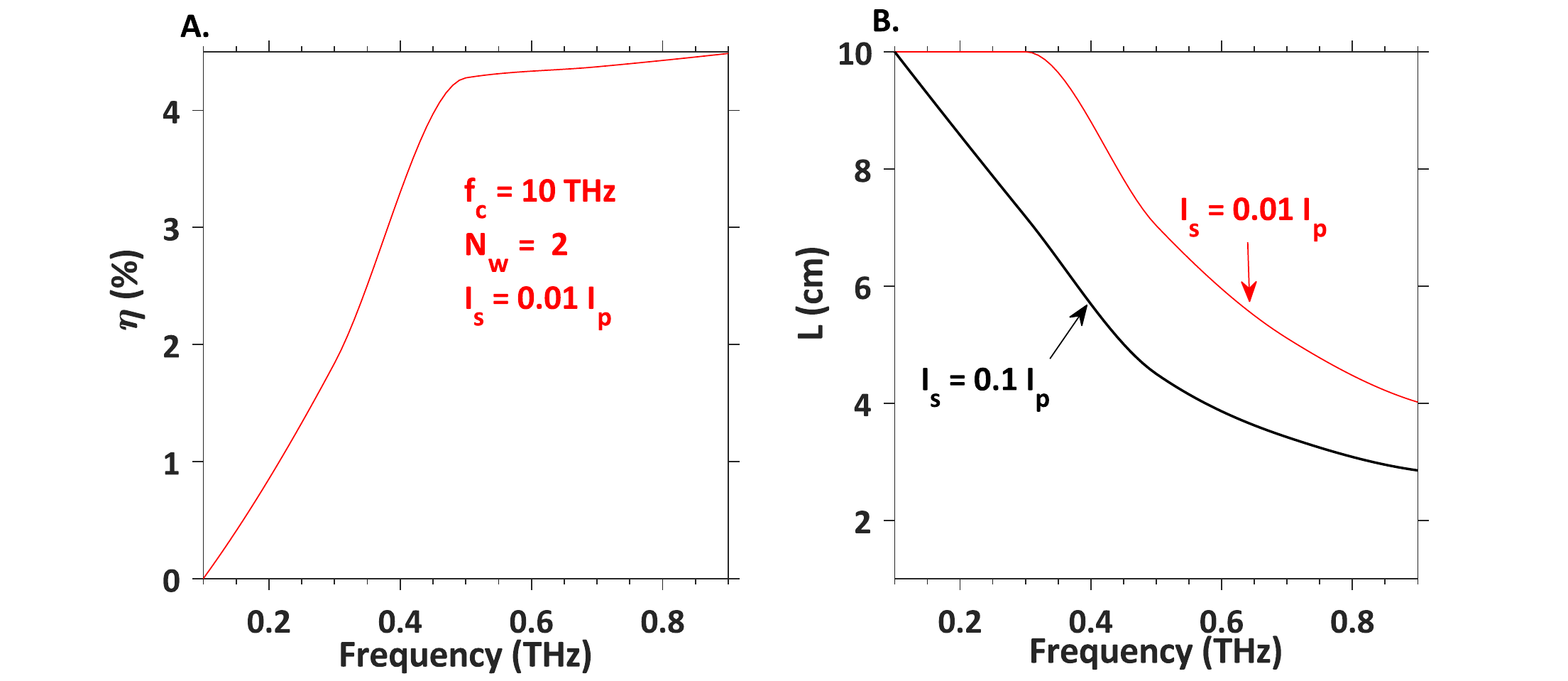}
\caption{\label{fig6}(a) Terahertz conversion efficiency as function of terahertz frequency shows initial increase and subsequent saturation due to increased absorption. (b) Requisite crystal lengths for maximizing efficiency show inverse dependence with terahertz frequency and peak seed intensity.}
\end{figure}

The above understanding might be extended to other relevant parameters such as the pump intensity or terahertz frequency. We may anticipate that larger terahertz frequencies close the gap faster and would hence experience terahertz growth at shorter distances. Furthermore, since conversion efficiency increases with terahertz frequency, larger frequencies may be anticipated to be generated with larger conversion efficiencies. However, the increase is not incessant due to the effects of absorption which also increase with frequency. This is evident in Fig. \ref{fig6}(a), which depicts the maximum conversion efficiency for various phase-matched terahertz frequencies. The material properties are once again obtained from Table \ref{tab1}. The PPLN period in each case is defined by $\Lambda= c(n_{THz}(f_{THz})-n_g)^{-1}/f_{THz}$, where $n_{THz}$ is the terahertz phase index and $n_g$ is the optical group refractive index. In these simulations, $f_c=10\,$THz while $N_w=2$. The peak intensity of the seed is set to be $I_s=0.01I_p$ and the total fluence to $0.7\,\text{Jcm}^{-2}$.

An increase in conversion efficiency with  terahertz frequency followed by saturation at $\approx 0.5\,$THz is observed in Fig. \ref{fig6}(a). Furthermore as anticipated, the optimal lengths reduce for larger phase-matched terahertz frequencies. A larger seed or pump intensity would expedite the process and also result in shorter interaction lengths (red curve Fig. \ref{fig6}(b)).

\subsection{Temporal evolution and damage thresholds}

In this section, we examine the temporal evolution of the total optical field, i.e. of both pump and seed put together. This provides guidelines for optimizing the input spectrum to mitigate laser-induced damage.

\begin{figure}
\centering
\includegraphics[scale=0.275]{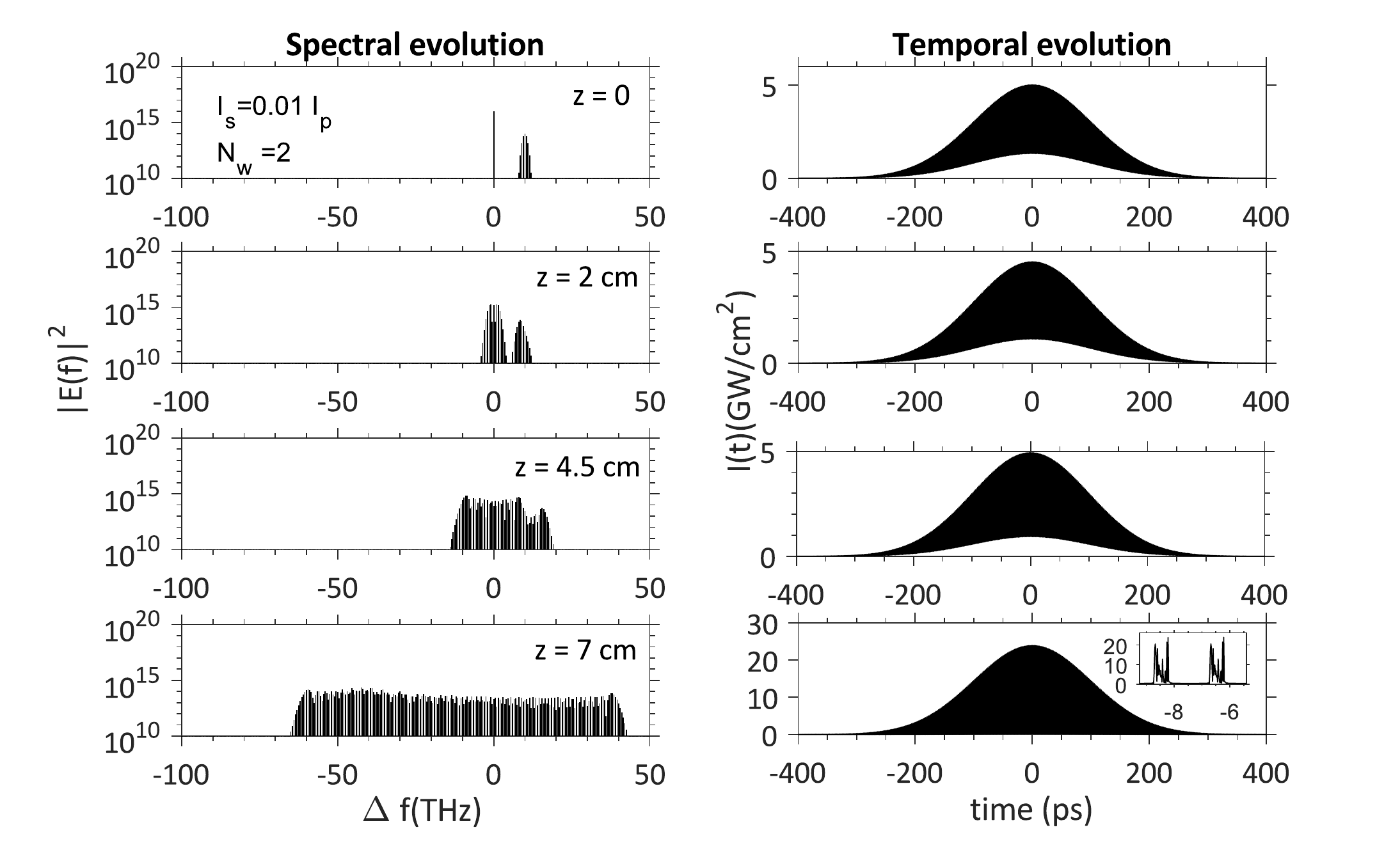}
\includegraphics[scale=0.33]{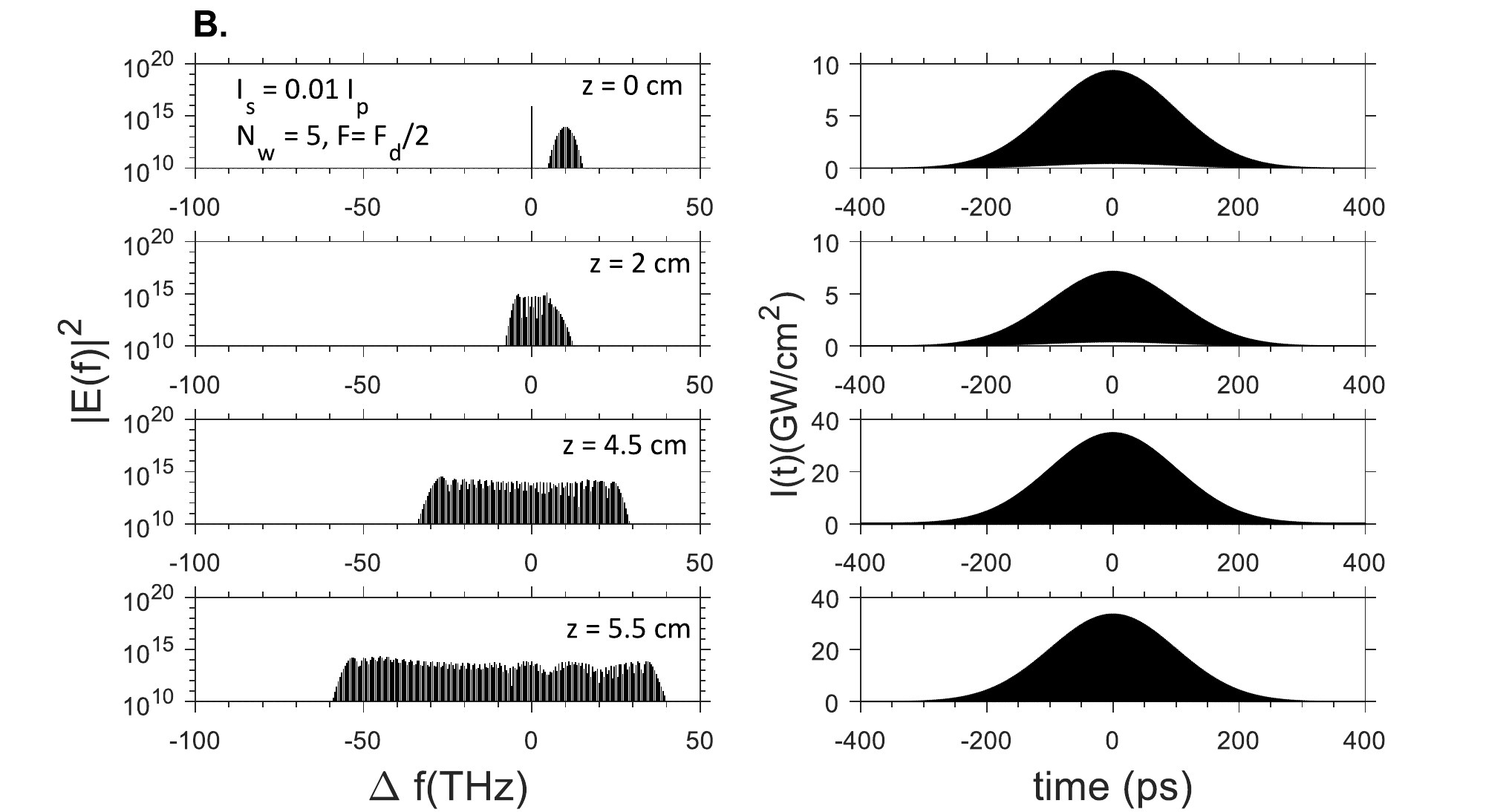}
\includegraphics[scale=0.3]{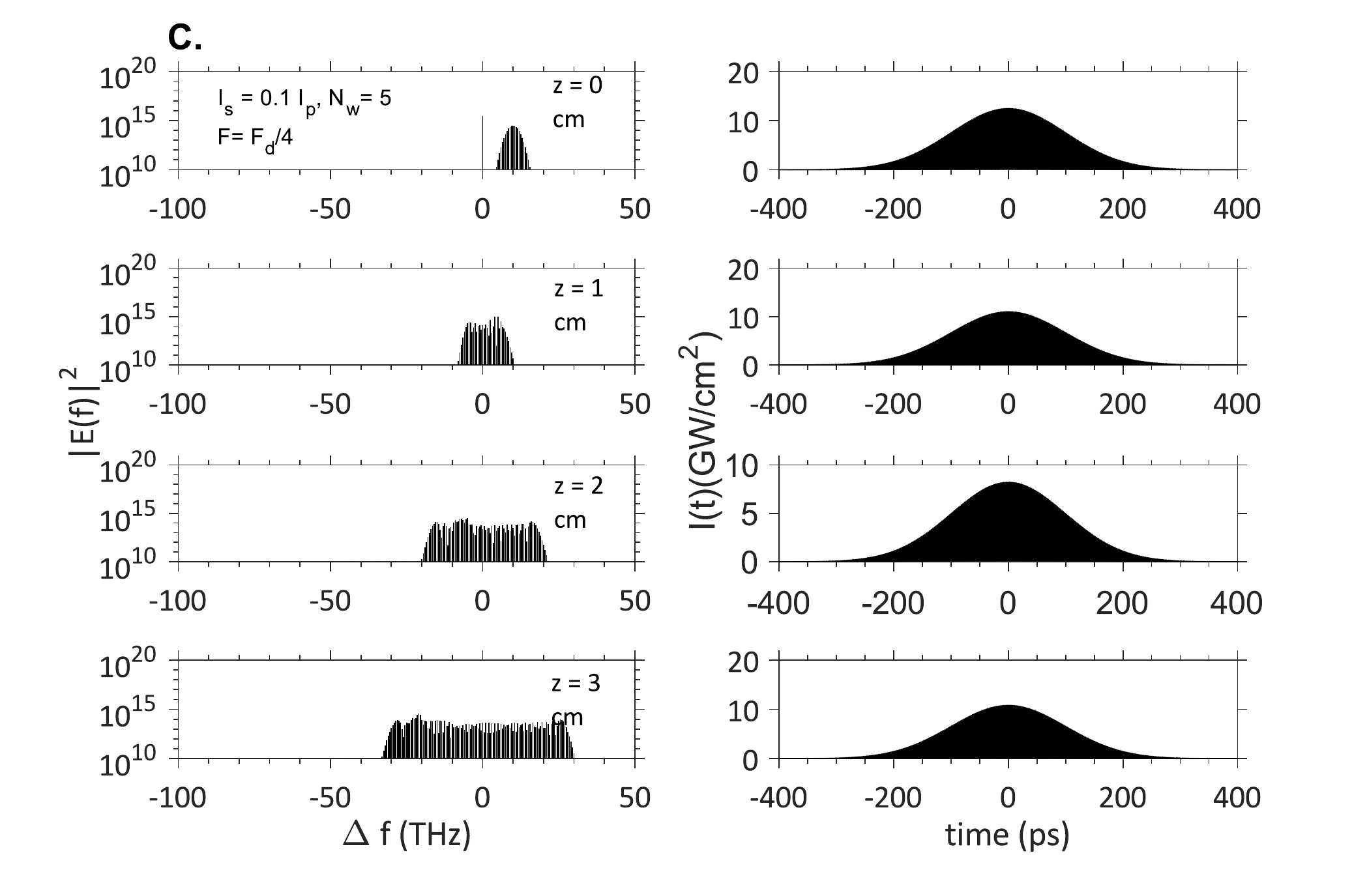}
\caption{\label{fig8} Spectral (left hand side panels) and temporal evolution (right hand side panels) along crystal length $z$ for various input pulse formats in (a)-(c). A larger initial bandwidth of seed (i.e. larger $N_w$) and larger ratio of $I_s/I_p$ enables keeping the peak intensity stable during the cascading process. However, while intensity growth is reduced in such cases, absolute intensity may still be higher as seen in (b). To circumvent this, the \textit{total} input fluence can be reduced as in (c). The maximum permissible intensity within the crystal is ascertained by tracking the generated free-electron density in the subsequent section. }
\end{figure}

In Fig. \ref{fig8}, we plot the spectral and temporal evolution of various input spectra along the length of the crystal. The spectral and temporal evolution up until the point when maximum optical intensity is reached along the crystal length is depicted. The temporal envelope of the total optical spectrum is given by $E(t,z) =\textbf{Re}\lbrace\Sigma_m A_m(z)e^{-jk_mz}e^{-t^2/\tau^2}\rbrace$, where $A_m$ corresponds to the $m^{th}$ optical line whose evolution is delineated in Eq. (\ref{op}). The panels on the left of Fig. \ref{fig8} represent the total optical spectrum at various values of $z$, while the right hand panels represent the corresponding intensity profiles $I(t)$ in the time-domain. 

In Fig. \ref{fig8}(a), we plot the spectral and temporal evolution for the pulse format utilized in Fig. \ref{fig3}(a) and \ref{fig4}(a). It is comprised of a pump separated by $f_c=10\,$THz from the seed with bandwidth $N_w=2$ in Eq. (\ref{seed_dist}). The peak seed intensity is given by $I_s=0.01I_p$. The total fluence of pump and seed taken together is set to half the laser-induced damage threshold value, i.e.  $F_d/2 = 0.7\,\text{Jcm}^{-2}$. A PPLN crystal phase-matched for the terahertz frequency $f_{THz}=0.5\,$THz at $T=100\,$K is assumed. Material and other properties are obtained from Table \ref{tab1}.  

As evident from the first panel on the right hand side of Fig. \ref{fig8}(a), at $z=0$, the combination of pump and seed represent an envelope with $e^{-2}$ duration $\tau=200$\,ps. The envelope is modulated at a rate of $0.5$\,THz or with a temporal period of 2\,ps. The total optical spectral bandwidth increases along the crystal length due to  cascading effects as evident in the left hand side panels of Fig. \ref{fig8}(a). In the time-domain, the modulations get deeper and eventually a pulse train with a period of 2\,ps is obtained (see inset of final panel on the right hand side of Fig. \ref{fig8}(a)). Importantly, one notices a dramatic increase in peak intensity to $>25\,\text{GWcm}^{-2}$ compared to the initial peak intensity of $5\,\text{GWcm}^{-2}$ which is an issue from the point of view of laser-induced damage. 

Since the intensity spiking behavior arises from a large increase in spectral bandwidth, it is presumable that starting with a larger number of lines around the seed centered at angular frequency $\omega_s$ would result in reduced intensity spiking. This is seen in Fig. \ref{fig8}(b), where $N_w=5$ and $I_s=0.01I_p$. As in Fig. \ref{fig8}(a), the total fluence is maintained at half the laser-induced damage threshold value of $F_d/2=0.7\,\text{Jcm}^{-2}$. It is evident in contrasting the first and last panels on the right-hand side of Fig. \ref{fig8}(b) that the intensity increase is only threefold as opposed to being more than fivefold in Fig. \ref{fig8}(a). However, the peak intensity at $z=5.5\,$cm in the last panel on the right hand side of Fig. \ref{fig8}(b) is $\sim 30,\text{GWcm}^{-2}$ which is larger than the value of $25,\text{GWcm}^{-2}$ in Fig. \ref{fig8}(a). This can be remedied as shown in Fig. \ref{fig8}(c).

In Fig. \ref{fig8}(c), we increase the fraction of the seed to pump intensity, i.e. $I_s/I_p$ to being $10\%$ rather than $1\%$. However, the total input fluence has to be reduced to maintain the peak intensity below reasonable levels. This is achieved by setting the total fluence to $F_d/4=0.35\,\text{Jcm}^{-2}$, which is half that in Figs. \ref{fig8}(a),(b). We find that in this case, even as cascading occurs,  the intensity level is roughly maintained at $10\,\text{GWcm}^{-2}$ in the right hand panels of Fig. \ref{fig8}(c). To ascertain exactly what peak intensity is permissible, we track free-electron densities in the subsequent section. 

\subsection{Free-electron density}

To determine what might be considered a reasonable input fluence, we use the generated free-electron density in the material as an indicator of the onset of laser-induced damage. This approach assumes the widely used  avalanche-breakdown model for laser-induced damage \cite{boyd2003nonlinear}. The free-electron density $N_c(t,z)$ is obtained according to Eq. (\ref{nc}).

\begin{gather}
\frac{dN_c(t,z)}{dt} = g_{av}I(t,z)N_c(t)+ E_g^{-1}\beta^{(4)}I^4(t,z) \label{nc}
\end{gather} 

In Eq. (\ref{nc}), the first term represents the generation of free electrons by an avalanche process. Here, $g_{av}$ is the avalanche coefficient while $I(t)$ is the temporal intensity profile corresponding to the combined pump and seed spectra. The second term in Eq. (\ref{nc}) corresponds to the generation of free electrons by four-photon absorption of photons in the $\approx 1\,\mu$m wavelength region. Here, $E_g$ is the bandgap of the material, while $\beta^{(4)}$ is the four-photon absorption coefficient. 

For lithium niobate and input radiation in the $1\,\mu$m region, a minimum of four photons would need to be absorbed since the bandgap of lithium niobate is $ E_g\approx 4\,$eV. However, for input radiation centered around  $800\,$nm, even three-photon absorption would have to be considered. Higher-order absorption coefficients are neglected since these events are of much lower probability for the typical intensity levels characterizing long pulses considered in the current system.

\begin{figure}
\centering
\includegraphics[scale=0.35]{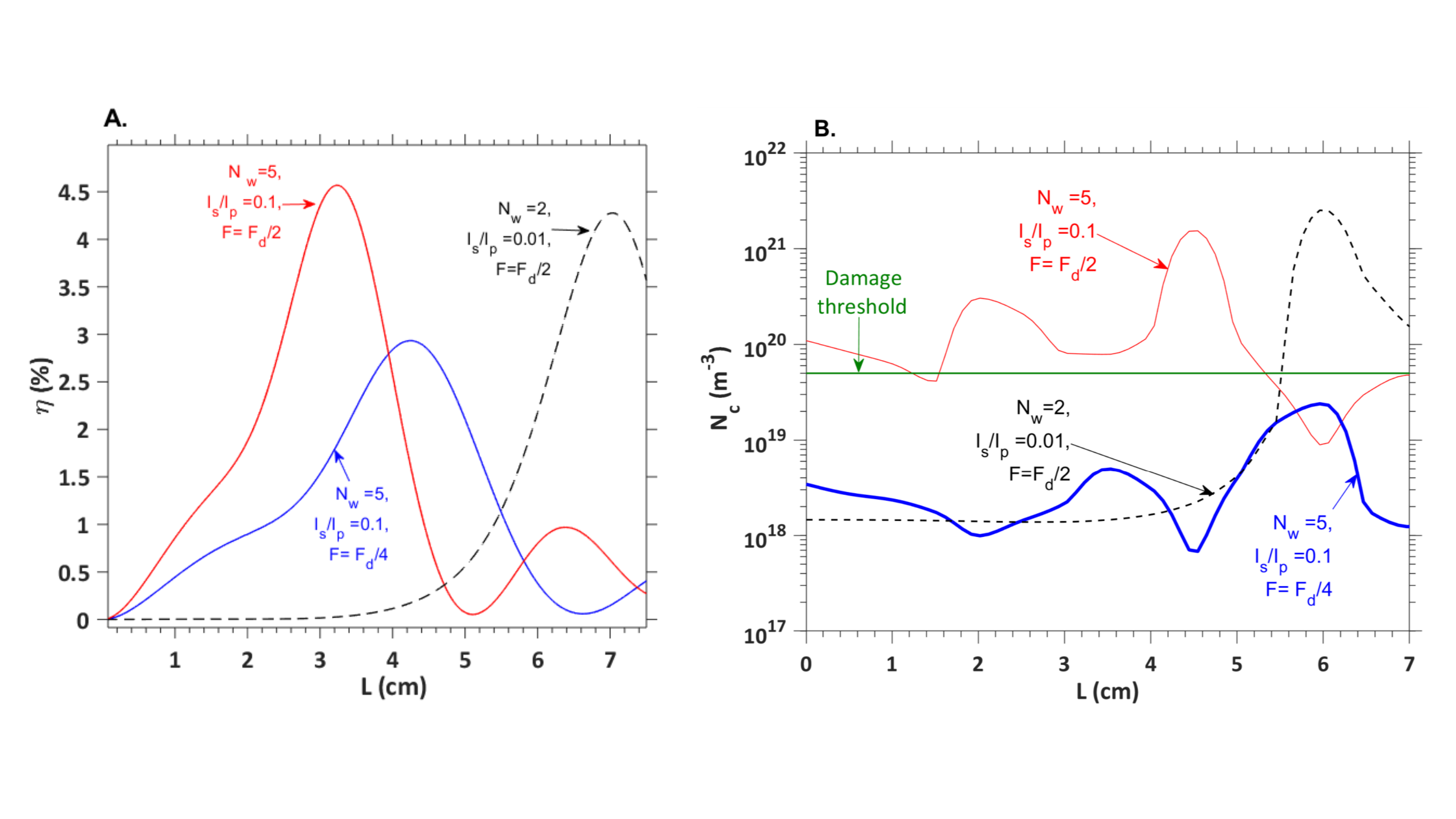}
\caption{\label{fig7}(a)Terahertz efficiency versus length for various values of $I_s/I_p$, number of seed lines $N_w$ and total fluence levels. (b) Corresponding carrier density generated for each case as a function of length. The green horizontal line represents the laser-induced damage threshold free-electron density calculated by using the laser-induced damage threshold fluence levels from Table \ref{tab1}.}
\end{figure}

Using parameters from Table \ref{tab1}, we plot the optical-to-terahertz conversion efficiency growth as a function of  crystal length for various pulse formats in Fig. \ref{fig7} (a) along with the maximum free-electron density (over time) along crystal length in \ref{fig7}(b).

As can be seen, for $N_w=2$, $I_s/I_p=0.01$ and a total input fluence of $F_d/2$, there is an exponential growth in the conversion efficiency $\eta$ beyond $z\approx4\,$cm. The increase in conversion efficiency is known to be accompanied by an increase in bandwidth due to cascading from Fig. \ref{fig8}(a). This in turn results in a spiking of optical intensity thereby resulting in an exponential growth in free-electron density in Fig. \ref{fig7}(b) (black-dotted curve). The  free-electron density rises above the threshold value (delineated by the green horizontal line) significantly. The threshold free-electron density is evaluated using a single pulse with $e^{-2}$ pulse duration  $\tau=200\,$ps and fluence given by the laser-induced damage threshold value from Table \ref{tab1}.

Increasing the ratio of seed to pump intensity $I_s/I_p$ to $10\%$ and $N_w$ to 5 alleviates temporal intensity growth. This in turn, eliminates the drastic increase in free-electron density along the crystal length characterizing the $N_w=2, I_s=0.01I_p$ case. This is seen in the red curve of Fig. \ref{fig7}(b). However, the value of the free-electron density is larger than the threshold (green) even at the outset. 

If we reduce the total input fluence for the $N_w=5,I_s=0.1I_p$ case by a factor of 2 to $F_d/4$, we retain the stability of intensity but also keep the free-electron density below the threshold value throughout (blue curve in Fig. \ref{fig7}(b)). 

At the same time, the conversion efficiency reduces from approximately $\approx4.5\%$ (red-solid, Fig. \ref{fig7}(a)) to $\approx 3\%$ (blue, Fig. \ref{fig7}(a)). Therefore, one can still obtain conversion efficiencies between $3-4.5\%$ while addressing the intensity spiking behavior due to increased bandwidth of the total spectrum.

Thus, in this section we've established that having a larger initial seed bandwidth can solve the intensity spiking behavior but will possess a lower effective damage fluence by virtue of larger peak intensities at the outset. This can be mitigated by reducing the total input fluence, while still obtaining high conversion efficiency, albeit lower.

\section{Application : Terahertz generation}

In this section, we evaluate terahertz generation with the pump located at $\lambda_p=1064\,$nm and seed lines centered about $\lambda_s=1030\,$nm. A cryogenically-cooled PPLN crystal at $T=80\,$K is assumed. The phase-matched terahertz frequency is set to $0.3$\,THz due to its relevance to compact electron accelerator technology \cite{kartner2016}. 

The choice of $\lambda_p,\lambda_s$ is informed by their suitability to being generated by high energy Nd:YAG and Yb:YAG lasers respectively. The pump and seed lines are all assumed to correspond to pulses of $\tau=200\,$ps $e^{-2}$ duration without loss of generality. For longer durations of $\sim$ns, a reduced conversion efficiency would be obtained which can be mitigated by recycling the cascaded optical spectrum in multiple crystals \cite{wang2018}.

We perform simulations \textit{without} the discrete approximations of Eqs. (\ref{thz})-(\ref{op}) based on the model presented in \cite{ravi2016_ps} and parameters from Table \ref{tab1}. The key equations are as follows:

\begin{subequations}
\begin{gather}
    \frac{dA_{THz}(\Omega,z)}{dz}=\frac{-\alpha(\Omega)}{2}A_{THz}(\Omega,z) -\frac{j\Omega^2\chi^{(2)}(z)}{2k(\Omega)c^2}\int_0^{\infty}A_{op}(\omega+\Omega,z)A_{op}^{*}(\omega,z)e^{-j\big[k(\omega+\Omega)-k(\omega)-k(\Omega)\big]z}d\omega\label{cont_thz}\\
    \frac{dA_{op}(\omega,z)}{dz} = -\frac{j\omega^2\chi^{(2)}(z)}{2k(\omega)c^2}\bigg[\int_0^{\infty}A_{op}(\omega+\Omega,z)A_{THz}^{*}(\Omega,z)e^{-j\big[k(\omega+\Omega)-k(\omega)-k(\Omega)\big]z}d\Omega\nonumber\\ +\int_0^{\infty}A_{op}(\omega-\Omega,z)A_{THz}(\Omega,z)e^{-j\big[k(\omega-\Omega+k(\Omega)-k(\omega)\big]z}d\Omega \bigg]\nonumber\\
    -\frac{j\varepsilon_0\omega_0n(\omega_0)n_{2}}{2}\mathfrak{F}\big[|A_{op}(t)|^2A_{op}(t)\big]\label{cont_op}
\end{gather}
\end{subequations}

Equation (\ref{cont_thz}) is analogous to Eq. (\ref{thz}) but considers the continuous terahertz spectrum $A_{THz}(\Omega,z)$. Notice that the second term in Eq. (\ref{cont_thz}) represents the ensemble of all DFG processes and is an integral representation of the second term in Eq. (\ref{thz}). Similarly, Eq. (\ref{cont_op}) is analogous to Eq. (\ref{op}). Here, $A_{op}(\omega,z)$ is used to encompass the complete optical spectrum. The first two terms of Eq. (\ref{cont_thz}) are analogous to the first two terms of Eq. (\ref{op}). The third term in Eq. (\ref{cont_op}) represents self-phase modulation with nonlinear refractive index $n_2$. For a more detailed explanation of Eqs. (\ref{cont_thz})-(\ref{cont_op}), the reader is referred to \cite{ravi2016_ps}. 

\begin{figure}
\centering
\includegraphics[scale=0.45]{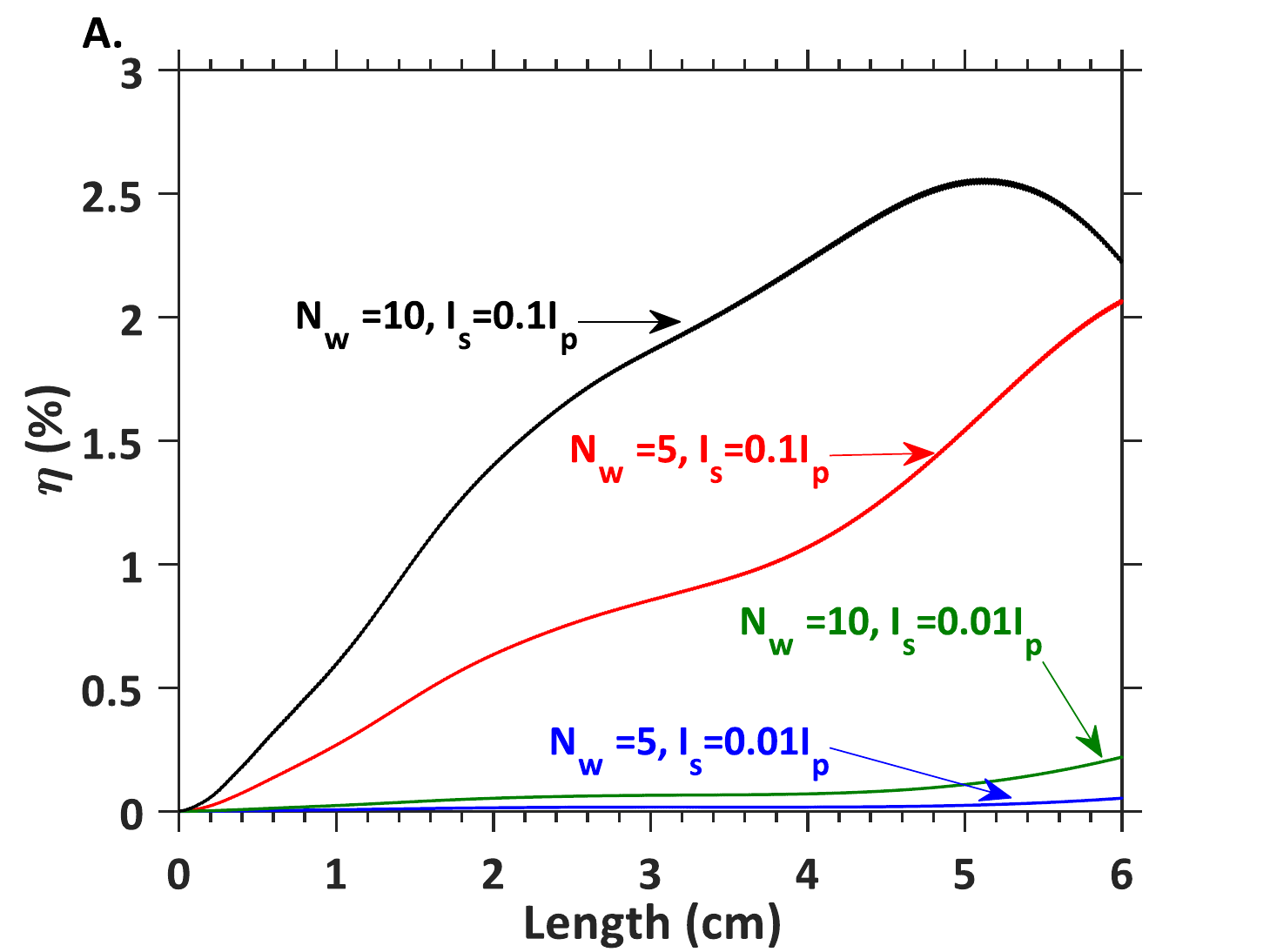}
\includegraphics[scale=0.4]{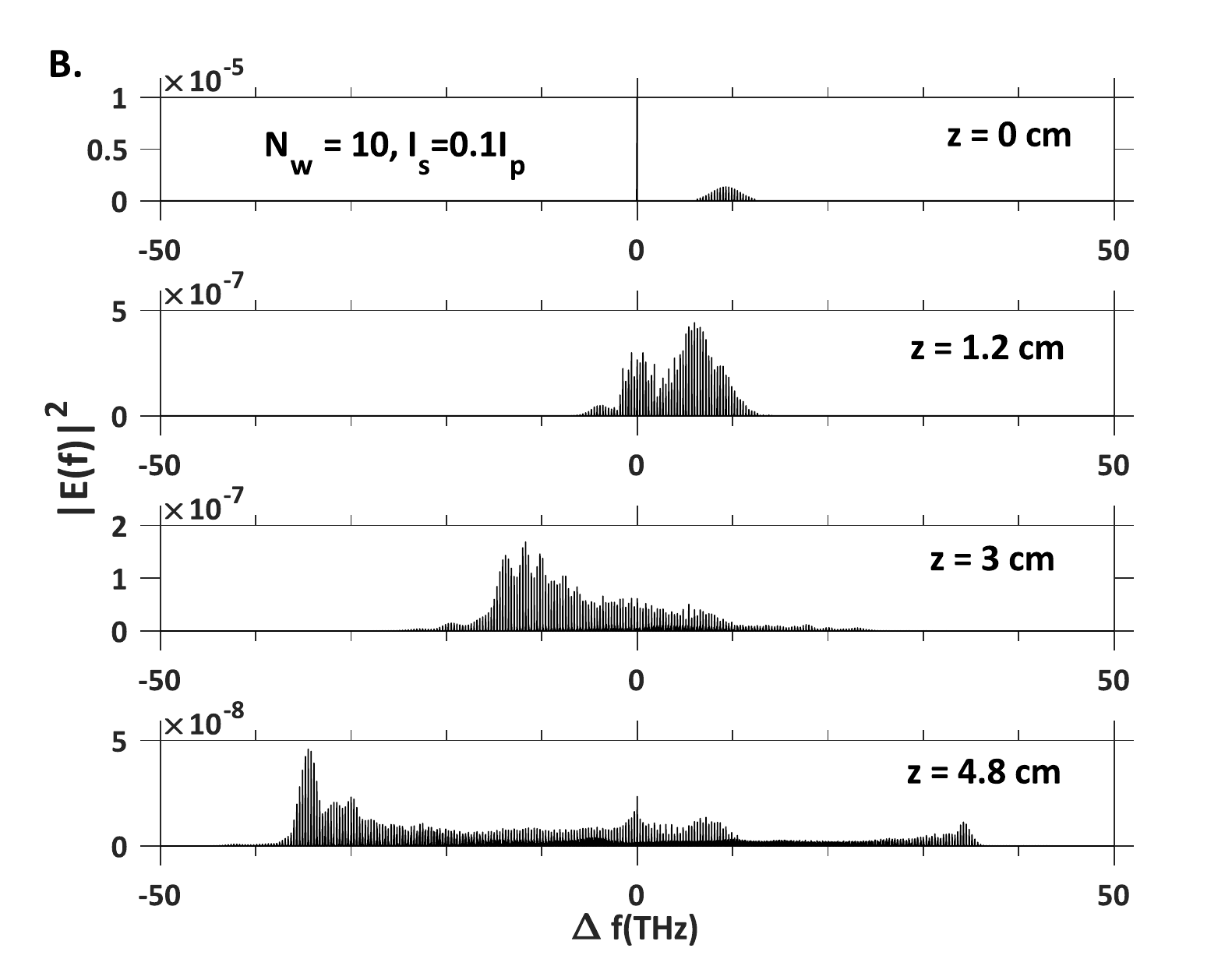}
\caption{\label{fig9}(a)Terahertz efficiency versus length in PPLN crystals phase matched for 0.3 THz  for various values of $I_s/I_p, N_w$. The pump is located at 1064\,nm and seed at 1030\,nm. (b) Spectral broadening and red-shift accompanying the percent-level conversion efficiencies.}
\end{figure}

In Fig. \ref{fig9}(a), the conversion efficiency for various values of $N_w$ and $I_s/I_p$ are presented. The total fluence is set to be $0.7\,\text{Jcm}^{-2}$, which is a factor of two lower than the laser-induced damage threshold fluence $F_d$ specified in Table \ref{tab1}. For cm$^2$-aperture PPLN crystals \cite{ishizuki2014}, this would permit pump energies of $>0.5\,$J. The optimization of fluence and input spectrum to mitigate laser-induced damage is not undertaken here for the sake of brevity. However, it would follow along the lines of section 3.6 and shall be reported elsewhere.

As described in section 3.4, the cascading process occurs more gradually with distance for lower terahertz frequencies. Therefore, in relation to the $0.5$ THz cases previously discussed, conversion efficiency growth for the 0.3 THz is slower with respect to crystal length. In Fig. \ref{fig9}(a), for $I_s/I_p=0.01$ the conversion efficiency is quite low for various $N_w$ even after 6\,cm of propagation. In principle, one may use multiple PPLN crystals and recycle the cascaded optical spectrum to circumvent this issue \cite{wang2018}.

However, if high efficiencies at the percent level need to be obtained at shorter crystal lengths, the ratio has to be increased to $I_s=0.1I_p$ as seen in Fig. \ref{fig9}(b).
In Fig. \ref{fig9}(a), for $N_w=10$ (black curve), the drop in efficiency is attributed to phase mismatch after the optical spectrum broadens significantly.

In Fig.\ref{fig9}(b), the optical spectrum at various $z$ locations for the case when $N_w=10,I_s=0.1I_p$ are presented. An increasing amount of red shift after merging of the pump and seed spectra, consistent with the previously developed physical picture is seen.

For the percent-level conversion efficiencies potentially achievable using these pulse formats and Joule-level pump energies, narrowband terahertz pulses with several millijoules of pulse energy thus appear feasible.

\section{Conclusion}

The spectral dynamics in the case of a system with pump and seed spectra separated by a large frequency offset is discussed when one of the pump or seed contains bandwidths larger than the terahertz frequency to be generated. In such a case, the phase-matched terahertz wave thus generated mediates the cascaded interaction between the pump and seed spectra. This can eventually result in the merging of the pump and seed spectra. For conditions of reducing phase-mismatch in the direction of the desired red-shift, high optical-to-terahertz conversion efficiency can be achieved. The issues of intensity spiking that accompany cascading effects was highlighted. Mitigating mechanisms using more broadband seed spectra was suggested. The approach was shown to be very valuable to generating narrowband terahertz radiation at percent-level conversion efficiencies and millijoule-level pulse energies using available 1064\,nm high energy lasers in combination with 1030\,nm seed pulses. The approach could be instrumental for enabling terahertz energy hungry applications such as electron acceleration and X-ray generation.\\

\textbf{Funding information}\\
This work was supported under by the Air Force Office of Scientific Research under grant AFOSR-A9550-12-1-0499, the European Research Council under the European Union's Seventh Framework Program (FP/2007-2013)/ ERC Grant Agreement n.609920 and the Center for Free-Electron Laser Science at DESY.\\

\bibliography{apssamp}

\end{document}